\documentclass{aa}  
\usepackage{graphicx}
\usepackage{txfonts}
\begin{document}

   \title{Magnetic braking at work in binaries}

   \author{W. Van Rensbergen and J.P. De Greve}

   \offprints{W. Van Rensbergen}

   \institute{Astrophysical Institute, Vrije Universiteit Brussel, Pleinlaan 2, 1050 Brussels, Belgium\\
   \email {wvanrens@vub.be}
   }

    \date{Received March 23, 2020}

 
  \abstract
   {Progenitors of binaries were determined with our binary evolutionary code so as to fit best with the presently observed positions of donor and gainer in the HR diagram. Moreover binaries with accretion disks around the gainer star were fitted towards the observed disk characteristics. At that stage magnetic braking was not included in the code. And no prediction of the rotation characteristics of the gainer were made.}
   {Equatorial velocities are measured for a number of mass gaining stars in interacting binaries. Tides tend to synchronize the rotation of the gainer but many observed low equatorial velocities can not be explained by tidal interaction alone.}
   {We introduced magnetic braking into our code so as the reproduce the observed equatorial velocities better.}
   {Large equatorial velocities of mass gaining stars are lowered by tidal interaction and magnetic braking. Tides are mainly at work at small orbital periods leaving magnetic braking alone at work at large orbital periods.}
   {Slow rotation is well reproduced by our code. But (not observed) critical rotation of the gainer in some systems can not be avoided by our calculations.}

   
    \authorrunning{W. Van Rensbergen et al.}  
     \titlerunning{Magnetic braking in binaries}
   \maketitle

\section{Introduction}
Van Rensbergen, De Greve et al. (\cite{VanRensbergenetal0}) published results of binary evolutionary calculations, including tidal interaction as proposed by Wellstein  (\cite{Wellstein}). We published a catalog of progenitors that produce the position of the two components of the binary in the HRD best. The evolution of the orbital period was determined using the law of conservation of angular momentum. The amount of angular momentum lost through stellar wind was calculated using Vink et al. (\cite{Vinketal}) for stars hotter than 12500 K and De Jager et al. (\cite{Dejageretal}) for cooler stars. In the case of liberal evolution the loss of angular momentum was calculated by Van Rensbergen et al.  (\cite{VanRensbergenetal1},\cite{VanRensbergenetal2}) assuming that mass lost from the system takes only the specific orbital angular momentum of the gainer. A detailed treatment of tides was proposed by Van Rensbergen \& De Greve (\cite{VanRensbergenetal}). Herein tides are very different for a star with a radiative atmosphere from one in convection (see e.g. Hilditch (\cite{Hilditch})). The convective mode was always applied during RLOF. Meridional circulation, as proposed by Tassoul (\cite{Tassoul}), was added as a significant contributor to the tidal action. Our paper modeled accretion disks characteristics around gainers.

\vspace{0.2cm}

The spin of the stars was not considered in these studies.

\vspace{0.2cm}

Observed equatorial velocities of gainer stars were published by Van Hamme \& Wilson (\cite{VanHammeWilson}), Miller et al.  (\cite{Milleretal}), Glazunova et al. (\cite{Glazunovaetal}) and  Dervisoglu et al. (\cite{Dervisogluetal}). The evaluation of the rotational velocities can only be explained introducing magnetic braking into the code. Without magnetic braking to many gainers would rotate at critical velocity.


\section{Generation of the magnetic field of the gainer}
\label{Magnetic_Field}

%

A solid rotator with constant angular velocity $\Omega$ from the center to the edge of the star can not develop a magnetic field.  In this paper we use magnetic fields that are produced by the dynamo of Spruit (\cite{Spruit}) which is at work when the angular velocity rises with an amount $\Delta\Omega$ over a distance $\Delta r$. The magnetic field is then directly proportional to q: 

\begin{equation}
q={{\Delta\Omega\over \Omega}\div{\Delta r\over r}}
\label{qSpruit}
 .\end{equation} 
 
In order to calculate q we introduce differential rotation into our model. We consider the gainer as a star that is composed as a core surrounded by a shell. Before the start of RLOF both parts rotate synchronously and produce no magnetic field. Magnetic fields can develop from the start of RLOF on.

\vspace{0.2cm}

Apart from the evaluation of the value of q in relation (\ref{qSpruit}), all the quantities needed to calculate the magnetic field with the dynamo of Spruit (\cite{Spruit}) are introduced in our binary evolutionary code at the interface between the gainer's core and shell. This means that the Brunt-V{\"{a}}s{\"{a}}l{\"{a}} frequency N, the thermal conductivity ${\kappa}$ and the magnetic diffusivity ${\eta}$ are evaluated in the gainer at every stage of the evolution of the binary.

\vspace{0.2cm}

Following Spruit (\cite{Spruit}), the thermal conductivity can be ignored (underscore (0)) when $B_{0}\over{B_{c0}}$ $<$ 1 so that the dynamo (underscore (1))  $B_{1}\over{B_{c1}}$ $>$ 1 creates the magnetic field. We followed this logic. But when both quotients are larger than 1, the mean value of $B_{r0}$ and $B_{r1}$ was taken for the magnetic field that governs the magnetic braking, using the following formulae:

\begin{eqnarray}
B_{{\phi}0}={r~{{(4~\pi}{\rho^{2}})^{{1}\over{2}}}~q~{{\Omega}^{2}}\over{N}}~;~{{B_{r0}}\over{B_{{\phi}0}}}=q~\left({{\Omega}\over{N}}\right)^{2}\\
B_{{\phi}1}=~r~{(4~\pi}{\rho^{2}})^{{1}\over{2}}~q^{{1}\over{2}}~ \Omega~{\left({{\Omega}\over{N}}\right)}^{{1}\over{8}}~{\left({{\kappa}\over{N~r^{2}}}\right)}^{{1}\over{8}}~;~{{B_{r1}}\over{B_{{\phi}1}}}={\left({{\Omega}\over{N}}\right)}^{{1}\over{4}}~{\left({{\kappa}\over{N~r^{2}}}\right)}^{{1}\over{4}}
\label{Dynamo}
\end{eqnarray}

The values in the right hand side of relations (2) and (3) are taken at the interface between the core and the shell.

\vspace{0.2cm}

From the start of RLOF on, the core is not influenced by the matter coming from the donor impinging on the gainer. The amount of angular momentum that is added only to the shell is given by Packet (\cite{Packet}), corrected with the impact-parameter ${d}\over {R_g}$:

\begin{equation}
\Delta J_{spin,shell}^{+}=6.04534 ~10^{54}~R_{g}~ \left(M_{g}+ {\Delta M_{g}\over2}\right)^{1\over2}~\left({d\over R_g}\right)
\label{upspin}
 .\end{equation}  

$\Delta J_{spin,shell}^{+}$ is expressed in cgs units, whereas masses and radii are respectively in $M_{\odot}$ and $R_{\odot}$.

\vspace{0.2cm}

With every value of $J_{spin,shell}$ corresponds a value of $\Omega_{shell}$=$J_{spin,shell} \over I_{shell}$ which is a characteristic value of the angular velocity in the shell. A magnetic field will be created when $\Omega_{shell}$ $>$ $\Omega_{core}$.
The radius dependent angular velocity $\Omega (r)$ raises continuously from $\Omega_{core}$ at the interface between core and shell to $\Omega_{edge}$ at the edge of the gainer. We assume that this raise follows the shape of an ellipse. In that case one obtains:

\begin{equation}
\Omega_{edge}= {4\over \pi}  \left(\Omega_{shell}-\Omega_{core}\right) + \Omega_{core}
\label{Edge}
 .\end{equation} 
  
\section{Conservation of angular momentum}
\label{sec_Conservation}

The total angular momentum  $J_{\Sigma}$ of a binary is the sum of the orbital angular momentum  $J_{orb }$ and the spin angular momentum of gainer  $J_{g }$ and donor $J_{d}$. Tides continuously exchange amounts of $\Delta J_{orb}$, $\Delta J_{g}$ and $\Delta J_{d}$. In the conservative case one has:

\begin{equation}
\Delta J_{orb} + \Delta J_{g} + \Delta J_{d} = 0
\label{ConservativeSum}
 .\end{equation} 

Angular momentum loss due to stellar wind (SW), mass loss from the system during liberal RLOF (OUT) and angular momentum loss due to magnetic braking (MAG) do not violate the law of conservation of angular momentum, but change it into:

 \begin{equation}
\Delta J_{orb} + \Delta J_{g} + \Delta J_{d} - \Delta J_{d+g}^{SW+OUT} - \Delta J_{g}^{MAG} = 0
\label{LiberalSum}
 .\end{equation}  

The values of  $\Delta J_{d+g}^{SW+OUT}$ and $J_{g}^{MAG}$ are negative. 
In this paper we only outline the effect of magnetic braking, since this was not included in our previous papers.
The amount of angular momentum lost by the gainer due to magnetic braking is given by Dervisoglu et al. (\cite{Dervisogluetal}) in cgs units.

\begin{equation}
\Delta J_{g}^{MAG}=-\left(4.807043~10^{40}\left(dM\over dt\right)^{3\over 7}\left(R\right)^{24\over 7}\left(B\right)^{8\over 7}\left(M\right)^{-2\over 7}\Omega_{edge}\Delta t\right)
\label{Dervisoglu}
\end{equation} 
 
Herein the quantity B is the magnetic field yielded by the dynamo of Spruit (\cite{Spruit}). The quantity  $dM\over dt$ is stellar wind mass loss of the gainer in  $M_{\odot}\over year$. This wind carries  matter and angular momentum into space. R is the radius of the gainer in  $R_{\odot}$, B is the radial component of the magnetic field expressed in  Gauss, M is the mass of the gainer expressed in  $M_{\odot}$ and  $\Omega_{edge}$ is the angular velocity of the edge of the gainer, defined in relation (\ref{Edge}). So that $v_{eq,g}$=$\Omega_{edge}$~$R_{eq,g}$ .

\vspace{0.2cm}

Relation (\ref{Dervisoglu}) is the expression derived by Dervisoglu et al. (\cite{Dervisogluetal}) for a magnetic dipole, which is valid in this case.

\subsection{Calculating the orbital period}
\label{sec_orbitalperiod}

The orbital angular momentum was calculated from relation (\ref{LiberalSum}). From these values the orbital period is calculated for circular orbits in cgs units with;

\begin{equation}
J_{orb}=1.045064~10^{51}~P^{1\over3}~\frac{M_{d}~M_{g}}{\left(M_{d}+M{g}\right)^{1\over3}}
\label{Period}
.\end{equation}
 
The orbital period P is in days and the masses in $M_{\odot}$.

\subsection{Extend of the shell}
\label{sec_AMFRAC}

The gainer is now composed by an inner core in solid rotation surrounded by a shell that rotates differentially. The core is not spun up nor braked down magnetically. Its rotation is only modulated by tides. The shell is spun up following relation (\ref{upspin}). Its rotation is modulated by tides and undergoes magnetic braking following relation (\ref{Dervisoglu}).

\vspace{0.2cm}

A core that takes X{\%} of the mass leaves (1-X) {\%} of the mass for the shell.

\vspace{0.2cm}

\begin{figure*}[!ht]
\centering
\includegraphics[width=9.6cm]{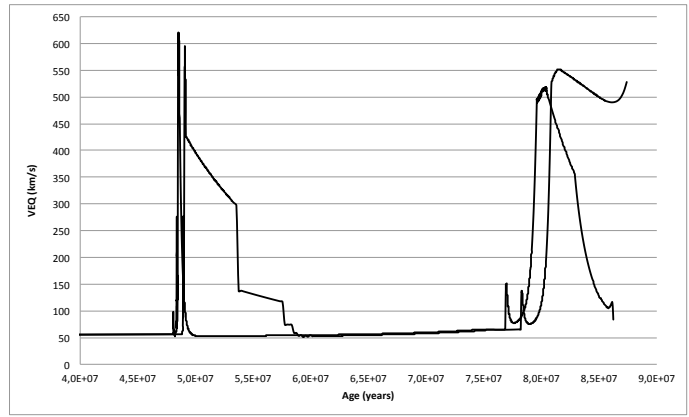}
\caption{Evolution of the equatorial velocity of the gainer of a $6.16 M_{\odot} + 2.9 M_{\odot}$ with an initial period of 1.6854 days, being a likely progenitor of  $\lambda$~Tau. In the upper curve tides act alone. In the lower curve tides and magnetic braking act together.}
\label{fig_fig1}
\end{figure*}

Figure (\ref{fig_fig1}) shows the evolution with time of the equatorial velocity of the gainer starting from a (6.16 + 2,9) binary with an initial orbital period of 1.6854~d. A possible progenitor for $\lambda$~Tau with initial value of X=0,95, undergoing tides and magnetic braking. This evolution with time is compared with the evolution of the gainer in the same system with a rigidly rotating and hence never magnetic gainer for which the rotation is thus modulated by tides only. This binary starts RLOF during core hydrogen burning of the gainer. Critical rotation of the gainer is achieved during the phase of rapid RLOF at around 48 million years after ZAMS. After that, RLOF will occur at a lower speed. Tidal interaction and magnetic braking will then be strong enough to synchronize the rotation of the gainer. In figure (\ref{fig_fig1}) one sees clearly that synchronization settles more rapidly when magnetic braking helps the tidal interaction. After around 65 million years RLOF will start again. Now during hydrogen shell burning of the donor. Critical rotation will again be achieved. But the rush to critical rotation is slowed down by the combined action of tides and magnetic braking. The present state of $\lambda$~Tau is in this up-going stage. The gainer of $\lambda$~Tau has there an equatorial velocity of 147 ${km \over sec}$, far below the critical value of 481 ${km \over sec}$. When finally critical rotation is achieved the orbital period is around 30 days and tides are then practically no more at work. Figure (\ref{fig_fig1}) shows that magnetic braking will prevent the gainer from critical rotation. This work is done at the very end of and after RLOF B.

\vspace{0.2cm}

\begin{figure*}[!ht]
\centering
\includegraphics[width=9.6cm]{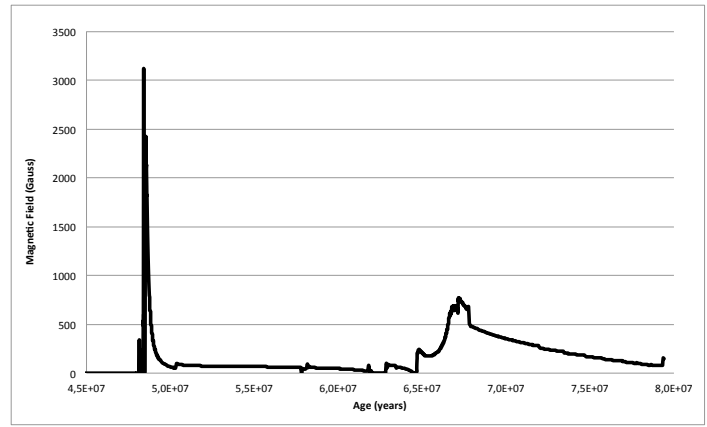}
\caption{Evolution of the magnetic field strength produced by the Spruit mechanism for the binary mentioned in Fig \ref{fig_fig1}}
\label{fig_fig2}
\end{figure*}

Figure (\ref{fig_fig2}) shows the evolution of the magnetic field of the gainer in the binary shown in  Figure (\ref{fig_fig1}). The magnetic field lives by the difference of angular velocity $\Omega$ between shell en core. Before RLOF A the gainer rotates at synchronous rotation. Shell and core, both at synchronous rotation. The core will always continue to rotate almost synchronously. When RLOF A starts, only the shell is spun up. When the shell rotates at critical velocity the magnetic field is at maximum (${\approx}$~3000~Gauss). When tides and magnetic braking have synchronized the rotation of the shell, the magnetic field disappears. From the beginning of RLOF B the shell is again spun-up and the magnetic field starts again to be build up. The value of the magnetic field is lower than during rapid RLOF A. A value of  ${\approx}$~750~Gauss is reached. Since magnetic braking is then nevertheless very active, the shell will gradually be synchronized. After that the magnetic field will again disappear.

\vspace{0.2cm}

Figure \ref{fig_fig2} shows the eras when the gainer can be regarded as a magnetic star.

\vspace{0.2cm}

It has to be mentioned that choices different from an initial value of X=0,95 (5\% of the mass of the gainer being in the shell), would hardly change the results shown in figures (\ref{fig_fig1})  and (\ref{fig_fig2}).

\vspace{0.2cm}

An initial shell with X=0.95 accretes  more and more mass and evolves finally into a massive shell with X=0,36 (64\% of the mass of the gainer being in the shell). This is not so different than the evolution with a shell with initial value of X=0.5 that evolves ultimately into a massive shell with X=0.2 (80\% of the mass of the gainer being in the shell), giving raise to magnetic fields and magnetic braking that are very similar to those obtained with an initial value of X=0,95. This value was used in all our calculations.

\vspace{0.2cm}

\begin{table*}
\begin{center}
\begin{tabular}{cccccccc} \hline
System - Progenitor & $M_{gainer}$ & $M_{donor}$ & $v_{eq}$ & R & Reference-Initial~Period~progenitor \\ \hline
$\beta$~Per & 3.70 & 0.81 & 51.50 & 0.00 & Dervisoglu et al.  (\cite{Dervisogluetal}) \\
3.41+1.1 & 3.69 & 0.82 & 48.77 & 0.00 & $P_{init}$= 1.146250\\
HS~Hya & 2.47 & 0.70 & 45.41 & 0.01 & Glazunova et al.  (\cite{Glazunovaetal}) \\
2.37+0.8 & 2.47 & 0.70 & 78.97 & 0.00 & $P_{init}$= 1.18912\\
CW~Eri & 2.59 & 0.74 & 33.28 & -0.01 & Glazunova et al.  (\cite{Glazunovaetal}) \\
2.23+1.1 & 2.59 & 0.74 & 41.92 & 0.01 & $P_{init}$= 1.30138\\
KO~Aql & 2.53 & 0.55 & 33.28 & -0.02 & Dervisoglu et al.  (\cite{Dervisogluetal})  \\
2.28+0.8 & 2.53 & 0.55 & 50.33 & 0.01 & $P_{init}$= 1.27150\\
ZZ~Boo & 3.43 & 0.96 & 9.51 & -0.02 & Glazunova et al.  (\cite{Glazunovaetal}) \\
2.59+1,8 & 3.49 & 0.90 & 48.19 & 0.01 & $P_{init}$= 1.5\\
Y~Psc & 2.80 & 0.70 & 38.05 & 0.00 & Dervisoglu et al.  (\cite{Dervisogluetal}) \\
2.3+1,2 & 2.80 & 0.70 & 47.58 & 0.03 & $P_{init}$= 1.34865\\
WW~Cyg & 2.10 & 0.60 & 41.01 & 0.03 & Dervisoglu et al.  (\cite{Dervisogluetal}) \\
1.5+1.2 & 2.10 & 0.60 & 51.35 & 0.06 & $P_{init}$= 1.138\\
AU~Mon & 5.93 & 1.18 & 126.32 & 0.25 & Dervisoglu et al.  (\cite{Dervisogluetal}) \\
4.16+3.00 & 5.93 & 1.19 & 104.87 & 0.22 & $P_{init}$= 2.005\\
AU~Mon & 5.97 & 1.19 & 218.01 & 0.54 & Van Hamme \& Wilson  (\cite{VanHammeWilson}) \\
4.07+3.09 & 5.96 & 1.19 & 322.85 & 0.64 & $P_{init}$= 2.003\\
V505~Sgr & 2.68 & 1.23 & 102.56 & 0.05 & Dervisoglu et al.  (\cite{Dervisogluetal}) \\
2.71+1.2 & 2.67 & 1.24 & 107.85 & 0.08 & $P_{init}$= 1.23198\\
TX~UMa & 4.76 & 1.18 & 63.62 & 0.04 & Dervisoglu et al.  (\cite{Dervisogluetal}) \\
4.24+1.7 & 4.75 & 1.19 & 71.78 & 0.00 & $P_{init}$= 1.44948\\
SZ~Psc & 3.00 & 0.77 & 9.26 & -0.03 & Glazunova et al.  (\cite{Glazunovaetal}) \\
2.47+1.3 & 3.00 & 0.77 & 44.38 & 0.03 & $P_{init}$= 1.4765\\
X~Tri & 2.43 & 1.21 & 50.00 & -0.16 & Van Hamme \& Wilson  (\cite{VanHammeWilson})  \\
2.44+1.2 & 2.43 & 1.21 & 93.42 & 0.02 & $P_{init}$= 0.98383\\
XY~Cet & 5.30 & 0.94 & 84.05 & 0.07 & Glazunova et al.  (\cite{Glazunovaetal}) \\
5.04+1.2 & 5.09 & 1.14 & 72.46 & 0.02 & $P_{init}$= 1.55429\\
RZ~Cas & 2.10 & 0.74 & 87.65 & 0.06 & Dervisoglu et al.  (\cite{Dervisogluetal})  \\
2.14+0.7 & 2.10 & 0.74 & 62.93 & 0.00 & $P_{init}$= 1.33437\\
U~Cep & 3.57 & 1.86 & 437.37 & 0.87 & Dervisoglu et al. (\cite{Dervisogluetal})  \\
3.33+2.1 & 3.56 & 1.87 & 488.95 & 0.95 & $P_{init}$= 2.13447\\
U~Cep & 4.41 & 2.83 & 280.17 & 0.66 & Van Hamme \& Wilson  (\cite{VanHammeWilson})  \\
4.24+3.0 & 4.39 & 2.85 & 463.03 & 0.96 & $P_{init}$= 2.35478\\
UV~Psc & 1.86 & 0.77 & 70.81 & 0.01 & Glazunova et al.  (\cite{Glazunovaetal}) \\
2.03+0.6 & 1.86 & 0.77 & 105.78 & -0.09 & $P_{init}$= 1.3999\\
AI~Dra & 2.37 & 1.09 & 86.90 & -0.06 & Van Hamme \& Wilson  (\cite{VanHammeWilson}) \\
2.36+1.1 & 2.36 & 1.10 & 85.71 & 0.04 & $P_{init}$= 1.18128\\
CD~Tau & 2.5 & 1.0 & 20.91 & 0.00 & Glazunova et al.  (\cite{Glazunovaetal}) \\
1.9+1.6 & 2.5 & 1.0 & 77.27 & 0.11 & $P_{init}$= 1.91047\\
AT~Peg & 2.50 & 1.21 & 84.51 & 0.01 & Dervisoglu et al. (\cite{Dervisogluetal})  \\
2.61+1.1 & 2.49 & 1.22 & 121.86 & 0.12 & $P_{init}$= 1.3406\\
TV~Cas & 3.78 & 1.53 & 80.48 & -0.03 & Dervisoglu et al. (\cite{Dervisogluetal})  \\
3.22+2.1 & 3.77 & 1.54 & 117.86 & 0.00 & $P_{init}$= 1.14467\\
$\lambda$~Tau & 7.19 & 1.87 & 90.69 & 0.05 & Van Hamme \& Wilson  (\cite{VanHammeWilson})  \\
6.16+2.9 & 7.15 & 1.87 & 146.73 & 0.20 & $P_{init}$= 1.68538\\
RW~Tau & 2.43 & 0.55 & 94.00 & 0.18 & Van Hamme \& Wilson  (\cite{VanHammeWilson})  \\
2.18+0.8 & 2.43 & 0.55 & 50.13 & 0.02 & $P_{init}$= 1.24613\\
\hline
\end{tabular}
\caption{Calculated velocities of gainers that fit observations best. Masses are in $M_{\odot}$, orbital periods in days and $v_{eq}$ in ${km \over sec}$. The masses mentioned by the observations of Gluenova et al. (\cite{Glazunovaetal}) are from Budding et al. (\cite{Buddingetal})}
\label{tab_tab1}
\end{center}
\end{table*}

\begin{table*}
\begin{center}
\begin{tabular}{cccccccc} \hline
System - Progenitor & $M_{gainer}$ & $M_{donor}$ & $v_{eq}$ & R & Reference-Initial~Period~progenitor \\ \hline
VZ~Hya & 2.52 & 0.89 & 19.90 & 0.00 & Glazunova et al.  (\cite{Glazunovaetal}) \\
2.01+1.4 & 2.52 & 0.89 & 107.24 & 0.17 & $P_{init}$= 1.47042\\
Z~Vul & 5.39 & 2.26 & 135.02 & 0.18 & Van Hamme \& Wilson  (\cite{VanHammeWilson}) \\
5.65+2.0 & 5.36 & 2.28 & 142.71 & 0.00 & $P_{init}$= 3.07536\\
IM~Aur & 2.38 & 0.77 & 139,76 & 0.20 & Van Hamme \& Wilson  (\cite{VanHammeWilson}) \\
2.35+0.8 & 2.38 & 0.77 & 70.04 & 0.00 & $P_{init}$= 1.15531\\
DL~Vir & 2.18 & 1.06 & 121.00 & 0.20 & Van Hamme \& Wilson  (\cite{VanHammeWilson}) \\
2.44+0.8 & 2.17 & 1.07 & 136.81 & 0.00 & $P_{init}$= 2.18242\\
$\delta$~Lib & 4.76 & 1.67 & 68.85 & -0.09 & Van Hamme \& Wilson  (\cite{VanHammeWilson})  \\
3.93+2.5 & 4.75 & 1.69 & 113.16 & 0.12 & $P_{init}$= 1.23263\\
TW~Dra & 1.70 & 0.80 & 37.09 & -0.02 & Dervisoglu et al.  (\cite{Dervisogluetal}) \\
1.5+1.0 & 1.70 & 0.80 & 121.51 & 0.63 & $P_{init}$= 2.092\\
V356~Sgr & 10.40 & 2.80 & 212,81 & 0.37 & Van Hamme \& Wilson (\cite{VanHammeWilson}) \\
8.7+6 & 10.90 & 2.64 & 118.34 & 0.14 & $P_{init}$= 1.86560\\
RX~Gem & 4.40 & 0.80 & 157.60 & 0.38 & Dervisoglu et al.  (\cite{Dervisogluetal}) \\
3.0+2.2 & 4.40 & 0.80 & 298.59 & 0.69 & $P_{init}$= 1.85226\\
TW~And & 1.68 & 0.32 & 31.64 & 0.01 & Glazunova et al.  (\cite{Glazunovaetal}) \\
1.4+0.6 & 1.68 & 0.32 & 155.42 & 0.32 & $P_{init}$= 1.08976\\
W~Del & 2.01 & 0.42 & 30.00 & 0.03 & Van Hamme \& Wilson  (\cite{VanHammeWilson}) \\
1.53+0.9 & 2.01 & 0.42 & 169.80 & 0.36 & $P_{init}$= 1.10746\\
SW~Cyg & 2.50 & 0.50 & 197.47 & 0.46 & Dervisoglu et al.  (\cite{Dervisogluetal}) \\
2.1+0.9 & 2.50 & 0.50 & 73.25 & 0.13 & $P_{init}$= 1.32299\\
RY~Per & 6.24 & 1.69 & 214.60 & 0.39 & Dervisoglu et al.  (\cite{Dervisogluetal}) \\
4.45+3.40 & 6.22 & 1.63 & 556.23 & 0.99 & $P_{init}$= 1.98167 \\
RS~Cep & 2.83 & 0.41 & 170.23 & 0.33 & Dervisoglu et al.  (\cite{Dervisogluetal}) \\
2.04+1.2 & 2.83 & 0.41 & 412.17 & 0.99 & $P_{init}$= 1.32215 \\
TT~Hya & 2.77 & 0.63 & 168.90 & 0.33 & Miller et al.  (\cite{Milleretal}) \\
2.0+1.4 & 2.77 & 0.63 & 482.24 & 0.99 & $P_{init}$= 1.68341 \\
AD~Her & 2.90 & 0.90 & 143.79 & 0.31 & Dervisoglu et al.  (\cite{Dervisogluetal}) \\
2.7+1.1 & 2.90 & 0.91 & 393.16 & 0.99 & $P_{init}$= 6.682 \\
RY~Gem & 2.66 & 0.24 & 70.53 & 0.14 & Glazunova et al.  (\cite{Glazunovaetal}) \\
2.35+0.55 & 2.61 & 0.24 & 376.12 & 0.87 & $P_{init}$= 1.12077\\
TU~Mon & 12.6 & 2.7 & 153.02 & 0.18 & Dervisoglu et al.  (\cite{Dervisogluetal}) \\
11.5+4.3 & 12.09 & 2.74 & 621.52 & 0.98 & $P_{init}$= 1.75065 \\
RZ~Eri & 3.57 & 0.34 & 69.00 & 0.14 & Glazunova et al.  (\cite{Glazunovaetal}) \\
2.69+1.4 & 3.57 & 0.34 & 422.63 & 1.00 & $P_{init}$= 1.52436\\
U~Sge & 5.35 & 2.14 & 79.00 & 0.06 & Van Hamme \& Wilson  (\cite{VanHammeWilson})  \\
4.49+3.0 & 5.34 & 2.15 & 560.37 & 1.00 & $P_{init}$= 2.07585\\
U CrB & 6.78 & 2.87 & 60.59 & 0.04 & Van Hamme \& Wilson  (\cite{VanHammeWilson})  \\
5.25+4.4 & 6.76 & 2.88 & 533.20 & 0.99 & $P_{init}$= 2.06346\\
U CrB & 4.74 & 1.46 & 60.59 & 0.03 & Dervisoglu et al.  (\cite{Dervisogluetal}) \\
4.2+2 & 4.73 & 1.74 & 545.00 & 1.00 & $P_{init}$= 1.93043\\
RZ~Cnc & 3.20 & 0.54 & 25.92 & 0.01 & Glazunova et al.  (\cite{Glazunovaetal}) \\
2.44+1.3 & 3.30 & 0.44 & 519.43 & 1.00 & $P_{init}$= 3.49907\\
CQ~Aur & 2.98 & 0.61 & 14.80 & -0.02 & Glazunova et al.  (\cite{Glazunovaetal}) \\
1.99+1.6 & 2.98 & 0.61 & 450.48 & 0.99 & $P_{init}$= 1.98785\\
\hline
\end{tabular}
\caption{Calculated velocities of gainers that fit observations less good than in Table \ref{tab_tab1}. Masses are in $M_{\odot}$, orbital periods in days and $v_{eq}$ in ${km \over sec}$. The masses for V356Sgr are from Dominis et al.(\cite{Dominisetal}). The masses by the observations of Glazunova et al.  (\cite{Glazunovaetal}) are from Budding et al.  (\cite{Buddingetal}). V356~Sgr and TU~Mon experience a liberal era during their life.}
\label{tab_tab2}
\end{center}
\end{table*}

\vspace{0.2cm}

\section{Results}
\label{sec_Results}

Van Hamme and Wilson (\cite{VanHammeWilson}) define the quantity F=$\frac{v_{eq}}{v_{sync}}$ and R=$\frac{F_{gainer}-1}{F_{crit,gainer}-1}$.
R $\epsilon$ [0-1] is a measure of rotation. R is 0 at synchronous rotation and 1 at critical rotation.

\vspace{0.2cm}

Tables (\ref{tab_tab1}) \& (\ref{tab_tab2}) compare the equatorial velocities of the gainers as obtained by our code with the observations. The calculated numbers shown in these tables were taken at the presently observed orbital period. The ranking runs from perfect determination to bad determination of $v_{eq}$ from the top of Table (\ref{tab_tab1}) to the bottom of Table (\ref{tab_tab2}). A determination of $v_{eq}$ is considered to be good when $\Delta$R = ($R_{model}-R_{obs}$) is small (${\approx}$~0) and is unacceptable when the same quantity is large (${\approx}$1).

\begin{flushleft}
$\bullet$~Some equatorial velocities are well reproduced by our calculations. These results are shown in Table (\ref{tab_tab1})\\
$\bullet$~Few observed values of equatorial velocities far below the critical value, are calculated being critical when they occur during rapid RLOF. In this case the up-spinning can not be braked down by the combined action of tides and magnetic braking. These cases are shown at the end of Table (\ref{tab_tab2})\\
$\bullet$~Some observed values of equatorial velocities below synchronous velocity are usually not reproduced by our calculations since tides always tend to synchronize rotation\\
\end{flushleft}

\vspace{0.2cm}

\section{Conclusions}
\label{sec_Conclusions}

RLOF spins the gainer up to critical velocity. This critical rotation remains as long as the system lives if one does not account for tidal interaction and magnetic braking. A system that first has RLOF A during hydrogen core burning of the donor will have a gainer that will rotate critically. At short orbital periods tidal interaction will synchronize the rotation of the gainer. This synchronization will occur more rapidly due to magnetic braking. When RLOF A stops the rotation of the gainer will synchronize rapidly. A new regime of RLOF will soon after the exhaustion of hydrogen in the core, start with RLOF B during hydrogen shell burning of the donor. The gainer will again be spun up into critical rotation. The orbital period and hence separation of the stars in the binary will practically annihilate the tidal forces. Every braking will now be supplied by magnetic braking. Due to magnetic braking during RLOF B the rotation of the gainer will go slower into critical rotation. Once critical rotation is attained, magnetic braking will lower down the rotation from the very end of RLOF until the gainer leaves the main sequence.

\begin{acknowledgements}

{We thank prof. Henk Spruit for explaining clearly how his dynamo works}

 \end{acknowledgements}

\end{document}